\documentclass[onecolumn]{jpsj3}
\usepackage{txfonts}

\title{Important Roles of Te 5$p$ and Ir 5$d$ Spin-orbit Interactions on the Multi-band Electronic Structure
of Triangular Lattice Superconductor Ir$_{1-x}$Pt$_x$Te$_2$}

\author{
Daiki Ootsuki$^1$, Tatsuya Toriyama$^2$, Masakazu Kobayashi$^3$, Sunseng Pyon$^3$, Kazutaka Kudo$^3$, Minoru Nohara$^3$, Takuya Sugimoto$^1$, Teppei Yoshida$^4$,
Masafumi Horio$^5$, Atsushi Fujimori$^5$, Masashi Arita$^6$, Hiroaki Anzai$^6$, Hirofumi Namatame$^6$, Masaki Taniguchi$^{6,7}$, Naurang ~L.~Saini$^8$, Takehisa Konishi$^9$, 
Yukinori Ohta$^2$, and Takashi Mizokawa$^1$
}

\inst{
$^1$Department of Physics \& Department of Complexity Science and Engineering, University of Tokyo, 5-1-5 Kashiwanoha, Chiba 277-8561, Japan\\
$^2$Department of Physics, Chiba University, Inage-ku, Chiba 263-8522, Japan\\
$^3$Department of Physics, Okayama University, Kita-ku, Okayama 700-8530, Japan\\
$^4$Graduate School of Human and Environmental Studies, Kyoto University, Sakyo-ku, Kyoto 606-8501, Japan\\
$^5$Department of Physics, University of Tokyo, 7-3-1 Hongo, Tokyo 113-0033, Japan\\
$^6$Hiroshima Synchrotron Radiation Center, Hiroshima University, Higashi-hiroshima 739-0046, Japan\\
$^7$Graduate School of Science, Hiroshima University, Higashi-hiroshima 739-8526, Japan\\
$^8$Department of Physics, University of Roma "La Sapienza" Piazzale Aldo Moro 2, 00185 Roma, Italy\\
$^9$Graduate School of Advanced Integration Science, Chiba University, Inage-ku, Chiba 263-8522, Japan\\
}

\date{\today}

\abst{We report an angle-resolved photoemission spectroscopy (ARPES) study
on a triangular lattice superconductor Ir$_{1-x}$Pt$_{x}$Te$_2$
in which the Ir-Ir or Te-Te bond formation, the band Jahn-Teller effect,
and the spin-orbit interaction are cooperating and competing with one another. 
The Fermi surfaces of the substituted system are qualitatively similar to 
the band structure calculations for the undistorted IrTe$_2$ 
with an upward chemical potential shift due to electron doping. 
A combination of the ARPES and the band structure calculations indicates
that the Te $5p$ spin-orbit interaction removes the $p_x/p_y$ orbital degeneracy 
and induces $p_x \pm ip_y$ type spin-orbit coupling near the A point. 
The inner and outer Fermi surfaces are entangled by the Te $5p$ and Ir $5d$ 
spin-orbit interactions which may provide exotic superconductivity 
with singlet-triplet mixing.}

\kword{spin-orbit coupling, Ir$_{1-x}$Pt$_x$Te$_2$, orbital degeneracy, angle-resolved photoemission spectroscopy, \ldots}

\begin{document}
\maketitle

4$d$ and 5$d$ transition-metal compounds with $t_{2g}$ orbital degeneracy
exhibit rich and interesting physical properties due to the competition
and/or the collaboration between the spin-orbit interaction and the (band) 
Jahn-Teller effect.
For example, in perovskite-type Ca$_2$RuO$_4$, the Jahn-Teller splitting 
between the Ru 4$d$ $t_{2g}$ $yz/zx$ and $xy$ orbitals collaborates 
with the 4$d$ $t_{2g}$ spin-orbit coupling to stabilize the $yz+ixy$ or 
$zx+ixy$ orbital to give the magnetic anisotropy \cite{Nakatsuji2000,Mizokawa2001}.
In other perovskite-type Sr$_2$IrO$_4$, the strong spin-orbit interaction of 
the Ir 5$d$ $t_{2g}$ orbitals is dominating and provides the unique $j_{1/2}$ state 
\cite{Kim2008,Kim2009}.
On the other hand, in spinel-type CuIr$_2$S$_4$, the band Jahn-Teller effect 
and the Ir-Ir bond dimerization play essential roles in the charge-orbital ordering 
with the Ir 5$d$ $t_{2g}$ spin-orbit interaction being apparently inactive 
\cite{Nagata1994,Radaelli2002,Khomskii2005}.
The difference between the perovskite-type and spinel-type systems can be attributed
to that between the corner sharing and edge-sharing Ir$X_6$ octahedra.
Yet, in Na$_3$IrO$_3$ with the edge-sharing of Ir$X_6$ octahedra, whereas the $j_{1/2}$
state is suggested\cite{Jackeli2009,Shitade2009}, 
based on the {\it ab-initio} calculation, it has recently been proposed that 
the $j_{1/2}$ and $j_{3/2}$ states are mixed due to formation of Ir 5$d$ 
molecular orbitals \cite{Mazin2012}.
In this context, it is highly desirable to study the Ir 5$d$ $t_{2g}$ electronic
states in various Ir compounds in a systematic way.

The discovery of superconductivity in doped or intercalated IrTe$_2$
by Pyon {\it et al.} \cite{Pyon2012} and by Yang {\it et al.} 
\cite{Yang2012} have added a new family of Ir chalcogenides 
to the list of fascinating 4$d$ and 5$d$ electron systems.
IrTe$_2$ undergoes a structural phase transition at $\sim$ 270 K 
from the trigonal (P-3m1) to the monoclinic (C2/m) structure
\cite{Matsumoto1999}. 
IrTe$_2$ and its derivatives show an interesting interplay between
lattice instabilities and superconductivity in the triangular lattice. 
Since the Ir 5$d$-to-Te 5$p$ charge-transfer energy is found 
to be small \cite{Ootsuki2012}, the Te 5$p$ orbitals can play
significant roles as proposed by Fang {\it et al.} 
and Oh {\it et al.} \cite{Fang2012,Oh2013}
Also the multi-band electronic structure of the Ir 5$d$ and Te 5$p$ orbitals
in IrTe$_2$ can induce the (band) Jahn-Teller instability and 
the Peierls instability \cite{Ootsuki2013}. 
In the case of Ir$_{1-x}$Pt$_x$Te$_2$,
the Ir-Ir bond \cite{Pyon2012,Kiswandhi2013}, the Ir-Te bond, and/or the Te-Te bond 
\cite{Fang2012} can be affected by the Pt doing.
In addition, the large Ir 5$d$ spin-orbit interaction can entangle 
the spin and orbital degrees of freedom in IrTe$_2$ 
and may induce nontrivial topological states \cite{Yang2012}. 
In order to reveal possible roles of the Ir 5$d$ and Te 5$p$ spin-orbit 
interactions in the Ir$_{1-x}$Pt$_x$Te$_2$, we have performed angle-resolved 
photoemission spectroscopy (ARPES) of Ir$_{0.95}$Pt$_{0.05}$Te$_2$ and 
compared the ARPES results with theoretical calculations with and without 
the Ir 5$d$ and Te 5$p$ spin-orbit interactions.

Single crystal samples of Ir$_{1-x}$Pt$_{x}$Te$_2$ were prepared 
using a self-flux method \cite{Fang2012,Pyon2012b}.
The photoemission measurements were performed at beamline 9A, 
Hiroshima Synchrotron Radiation Center using a SCIENTA R4000 analyzer
with circularly polarized light. The total energy resolutions 
were set to 18 meV, 22 meV, and 29 meV for excitation energies 
of $h\nu$ = 23 eV, 26 eV, and 29 eV, respectively. 
The angular resolution was set to $\sim$ 0.3$^{\circ}$ that 
gives the momentum resolutions of $\sim$ 0.015 \AA$^{-1}$,
0.016 \AA$^{-1}$, and 0.017 \AA$^{-1}$ for $h\nu$ =23 eV, 
26 eV, and 29 eV, respectively. 
The incident beam is 50$^{\circ}$ off the sample surface. 
The base pressure of the spectrometer was in the $10^{-9}$ Pa range. 
The samples were cleaved at $20$ K under the ultrahigh vacuum. 
The samples were oriented by {\it ex situ} Laue measurements.
The spectra were acquired within 8 hours after the cleavage.
Binding energies were calibrated using the Fermi edge of
gold reference samples. 
For the band structure calculations, we employ the code WIEN2k \cite{Blaha2002} 
based on the full-potential linearized augmented-plane-wave method 
and present the calculated results obtained in the generalized gradient 
approximation (GGA) for electron correlations, where we use the 
exchange-correlation potential of Ref.21 \cite{Perdew1996}.  
The spin-orbit interaction is taken into account for both Ir and Te ions when necessary.  
We use the crystal structure and atomic positions measured for IrTe$_2$ at room temperature
\cite{Jobic1992}. 
In the self-consistent calculations, we use 264 $k$-points in the irreducible 
part of the Brillouin zone with an anisotropic sampling to achieve better convergence.  
Muffin-tin radii ($R_{\rm MT}$) of 2.50 (Ir) and 2.42 (Te) Bohr are used
and we assume the plane-wave cutoff of $K_{\rm max}=7.0/R_{MT}$.  

In Fig. 1, the ARPES spectra along the A-H direction 
of Ir$_{0.95}$Pt$_{0.05}$Te$_2$ are compared with those of IrTe$_2$ 
above the structural transition temperature. 
The ARPES spectra are compared with the band dispersions 
obtained for undistorted IrTe$_2$ using 
the GGA calculation with the Ir 5$d$ and Te 5$p$ spin-orbit interactions.
As for Ir$_{0.95}$Pt$_{0.05}$Te$_2$, it is assumed that
the Pt doping introduces 0.05 electron per Ir and that
the chemical potential is shifted upwards in the band 
structure for undistorted IrTe$_2$.
The experimental and theoretical band dispersions are in qualitative 
agreement for IrTe$_2$ above the transition temperature as well for 
Ir$_{0.95}$Pt$_{0.05}$Te$_2$. 
Near the Fermi level, as predicted by the calculation,
the inner band forms the small hole pockets around the A point 
(the inner Fermi surfaces) while the outer band forms the large Fermi 
surface (the outer Fermi surface).
If the Pt doping supplies electrons to the Ir 5$d$ and Te 5$p$ bands,
the entire band should be shifted downwards in going from
IrTe$_2$ to Ir$_{0.95}$Pt$_{0.05}$Te$_2$. 
Indeed, the calculation for Ir$_{0.95}$Pt$_{0.05}$Te$_2$ is 
shifted downwards assuming the rigid band model and agrees 
very well with the experimental result.
The energy shift between IrTe$_2$ and Ir$_{0.95}$Pt$_{0.05}$Te$_2$
is consistent with the rigid band shift of the calculated results,
indicating that the Pt doping provides electrons to the Ir 5$d$ 
and Te 5$p$ bands. 

Figure 2(a) and (b) show the Fermi surface maps for Ir$_{0.95}$Pt$_{0.05}$Te$_2$
taken at 23 eV with right-handed and left-handed circularly polarized light,
respectively. 
The six hole pockets around the A point (the inner Fermi surfaces) 
and the large Fermi surface (the outer Fermi surface) are partly 
observed as predicted by the band structure calculation.
As for the inner Fermi surfaces, the upper three hole pockets
tend to be emphasized with the right-handed circularly polarized light,
while the lower three hole pockets gain their intensity 
with the left-handed circularly polarized light.
The asymmetric intensity due to the transition-matrix element effect 
is partially removed by summing the Fermi surface maps \cite{Okazaki2012},
and the six fold symmetry of the six hole pockets are more clearly 
seen as displayed in Fig. 2(c). Here, the transition-matrix element 
effect still remains since the incident light is 50$^{\circ}$ off 
the sample surface.

Figures 3(a), (b) and (c) show the evolution of the Fermi surfaces by changing the photon energy.
The momentum perpendicular to the surface ($k_z$) is $\sim$ $\pi/c$, $9\pi/10c$, and $8\pi/10c$ 
for 23 eV, 26 eV, and 29 eV, respectively. 
The Fermi surface maps extracted from the ARPES data are compared with 
the theoretical Fermi surfaces which are obtained by the rigid band shift of the GGA calculation
with and without the spin-orbit interactions for undistorted IrTe$_2$ 
for $k_z$ = $\pi/c$, $9\pi/10c$, and $8\pi/10c$, respectively.
The calculations with and without the spin-orbit interactions 
indicates that the effect of the spin-orbit interaction is more
significant for the inner Fermi surfaces than the outer one.
At 23 eV, the area of the inner Fermi surfaces is well reproduced
by the calculation with the spin-orbit interactions as shown in Fig. 3(a).
The effect of the spin-orbit interaction is more pronounced at 29 eV.
Without the spin-orbit interactions, even the geometry of the 
calculated Fermi surfaces is different from the ARPES results.
On the other hand, the Fermi surfaces calculated with the spin-orbit interaction 
are consistent with the ARPES results.

In Fig. 3(d), the Fermi surface map is compared with the calculation
with (without) the Te 5$p$ (Ir 5$d$) spin-orbit interaction and 
that with (without) the Ir 5$d$ (Te 5$p$) spin-orbit interaction.
The calculation with the Ir 5$d$ spin-orbit interaction agrees
with the ARPES results than the calculation with the Te 5$p$ spin-orbit interaction,
indicating that the Ir 5$d$ spin-orbit interaction is more important 
than the Te 5$p$ spin-orbit interaction for the area of the inner Fermi surfaces.

The band dispersions along the A-H and A-L directions are compared with the calculated
results with and without the spin-orbit interactions in Figs. 4(a) and (b).
The spin-orbit band splitting at -0.5 eV between the inner bands around the A point 
and that at -0.25 eV between the inner and outer bands in the A-H cut
are clearly observed in the APRES results (indicated by the arrows),
as predicted by the GGA calculation with the spin-orbit interactions.
The GGA calculation predicts that, around the A point, the inner bands 
near the Fermi level are mainly derived from the Te 5$p_x$ and 5$p_y$ orbitals.
Therefore, the band splitting at -0.5 eV around the A point can be 
attributed to the $p_x \pm ip_y$ type spin-orbit coupling.
In going from the A point to the H point, the inner bands dominated
by the Te 5$p_x$ and 5$p_y$ orbitals tend to have some contribution
of the Ir 5$d$ $e_{g}^{\pi}$ orbitals. On the other hand,
the outer Fermi surface is mainly constructed from the Te 5$p_z$ orbitals
with some contribution of the Ir 5$d$ $a_{1g}$ orbitals.
The band splitting at -0.25 eV between the inner and outer bands in the A-H cut
would be due to the $p_z \pm ip_x$ type or $p_z \pm ip_y$ type Te 5$p$ spin-orbit 
coupling with some contribution of the Ir 5$d$ spin-orbit coupling.
In order to clarify the different roles of the Ir 5$d$ and Te 5$p$ 
spin-orbit interactions, the band dispersions along the A-H and A-L directions 
are compared with the calculation with (without) the Te 5$p$ (Ir 5$d$) 
spin-orbit interaction and that with (without) the Ir 5$d$ (Te 5$p$) 
spin-orbit interaction in Figs. 4(c) and (d). 
Although the intensity of the outer band is very small in Figs. 4(a) and (c), 
the dispersion of the outer band can be seen in the enlarged plot of Fig. 4(e).
The band splitting at -0.5 eV around the A point is dominated by   
the Te 5$p$ spin-orbit interaction which provides the $p_x \pm ip_y$ type 
spin-orbit coupling as discussed above. 
On the other hand, although the band splitting at -0.25 eV between the inner
and outer bands in the A-H cut is induced by the Te 5$p$ spin-orbit 
interaction, the magnitude of the splitting is much more enhanced 
by the inclusion of the Ir 5$d$ spin-orbit coupling. 
Therefore, as for the spin-orbit coupling between the inner and outer bands,
the coupling between the Ir 5$d$ $a_{1g}$ and Ir 5$d$ $e_{g}^{\pi}$ orbitals 
has a substantial contribution although the inner and outer bands are dominated
by the Te 5$p$ character.

In Ir$_{1-x}$Pt$_x$Te$_2$, the inner and outer bands are very close to each other 
near the Fermi level along the A-H direction, and the Te 5$p$ and Ir 5$d$ spin-orbit 
interactions introduce the strong entanglement between the spin and orbital parts 
of the electrons. The recent ARPES study on Sr$_2$RuO$_4$ revealed that
the strong spin-orbital entanglement provides mixing 
of spin-singlet and spin-triplet Cooper pairs and
may cause an exotic superconducting state \cite{Veenstra2013}.
Since the magnitude of the spin-orbit interaction in Ir$_{0.95}$Pt$_{0.05}$Te$_2$
is much stronger than that in Sr$_2$RuO$_4$, one can expect
stronger mixing between spin-singlet and spin-triplet Cooper pairing 
in the Ir$_{0.95}$Pt$_{0.05}$Te$_2$.

In conclusion, we have studied the multi-band electronic structure of
triangular lattice superconductor Ir$_{1-x}$Pt$_{x}$Te$_2$ using ARPES.
The Pt doping introduces the electrons to the Ir 5$d$ and Te 5$p$ bands
and the simple rigid band model is consistent with the observed band
dispersions and the Fermi surfaces. The inner and outer Fermi surfaces 
with the strong Te 5$p$ character are very close to each other 
along the A-H direction, and the strong Te 5$p$ and Ir 5$d$
spin-orbit interaction in this particular momentum region results in 
the mixing between the spin-singlet and spin-triplet Cooper pairs
in Ir$_{1-x}$Pt$_{x}$Te$_2$.

The authors would like to thank Profs. A. Damascelli and H.-J. Noh for valuable discussions.
This work was partially supported by Grants-in-Aid from the Japan Society of 
the Promotion of Science (JSPS) (22540363, 23740274, 24740238, 25400356) and 
the Funding Program for 
World-Leading Innovative R\&D on Science and Technology (FIRST Program) from JSPS.
T.T. and D.O. acknowledge supports from the JSPS Research Fellowship for Young Scientists. 
The synchrotron radiation experiment was performed with the approval of 
Hiroshima Synchrotron Radiation Center (Proposal No.13-A-6).

\begin{figure}
\includegraphics[width=10cm]{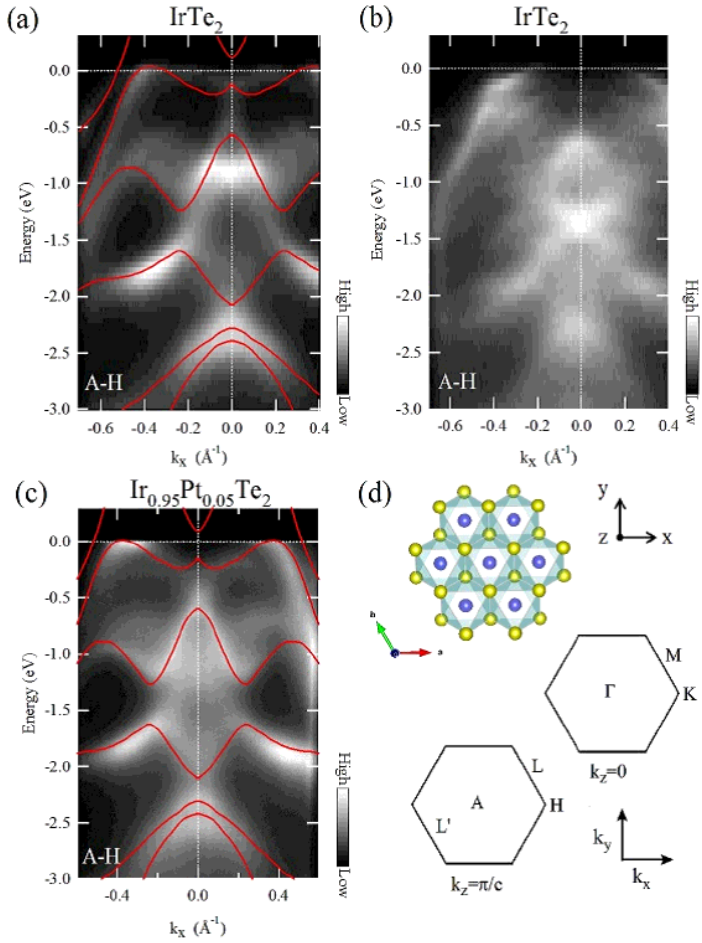}
\caption{(color online) 
(a) ARPES spectra of IrTe$_2$ along the A-H direction
taken at 300 K above the structural transition temperature.
The solid curves indicate the band dispersions 
obtained for undistorted IrTe$_2$ using 
the GGA calculation with the Te 5$p$ and Ir 5$d$ spin-orbit interactions.
(b) ARPES spectra of Ir$_{0.95}$Pt$_{0.05}$Te$_2$ along the A-H direction taken at 20 K. 
The solid curves indicate the calculation shifted by the electron doping by Pt. 
(c) Schematic drawings for the crystal structure visualized using 
the software VESTA \cite{Homma} and the first Brillouin zone of Ir$_{0.95}$Pt$_{0.05}$Te$_2$.}
\end{figure}

\begin{figure}
\includegraphics[width=9cm]{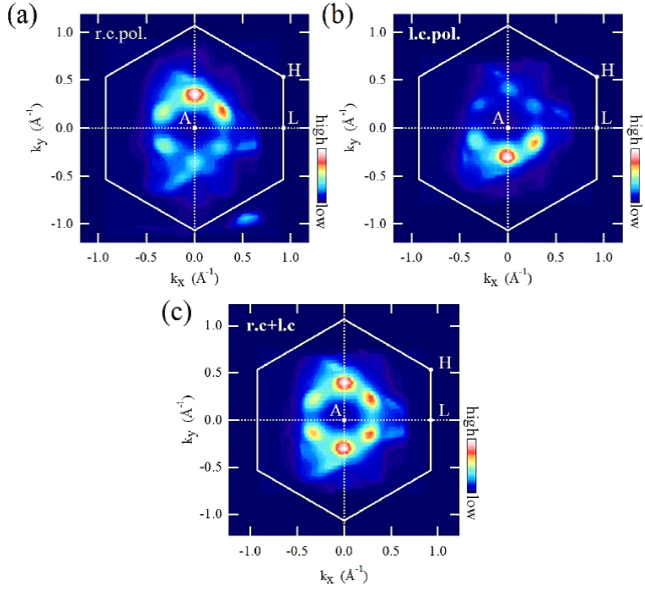}
\caption{(color online)
(a) Fermi surface map for Ir$_{0.95}$Pt$_{0.05}$Te$_2$
with right-handed circularly polarized light.
(b) Fermi surface map for Ir$_{0.95}$Pt$_{0.05}$Te$_2$
with left-handed circularly polarized light.
(c) Sum of the Fermi surface maps.
}
\end{figure}

\begin{figure}
\includegraphics[width=9cm]{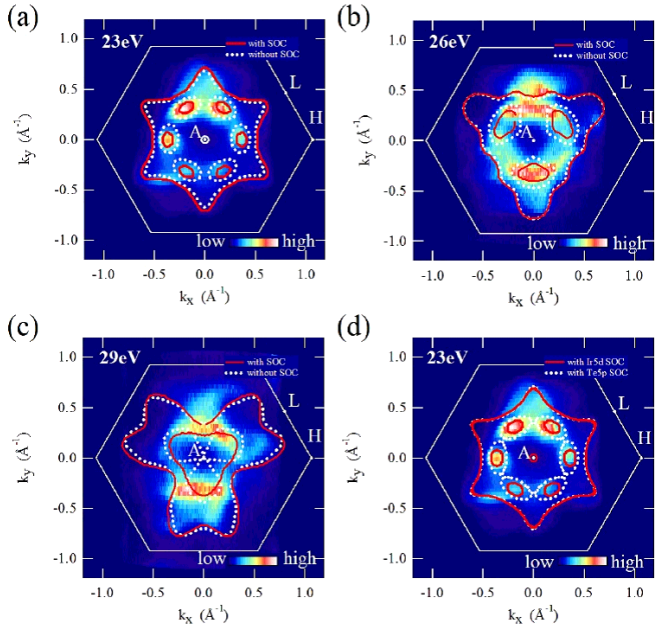}
\caption{(color online) 
(a) Fermi surface map for Ir$_{0.95}$Pt$_{0.05}$Te$_2$
at 23 eV compared with the GGA calculations with and without
the spin-orbit interactions.
(b) Fermi surface map for Ir$_{0.95}$Pt$_{0.05}$Te$_2$
at 26 eV compared with the GGA calculations with and without
the spin-orbit interactions.
(c) Fermi surface map for Ir$_{0.95}$Pt$_{0.05}$Te$_2$
at 29 eV compared with the GGA calculations with and without
the spin-orbit interactions.
(d) The GGA calculation with (without) the Te 5$p$ (Ir 5$d$) 
spin-orbit interaction and that with (without) the Ir 5$d$ (Te 5$p$)
comapred with the Fermi surface map at 23 eV.}
\end{figure}

\begin{figure}
\includegraphics[width=9cm]{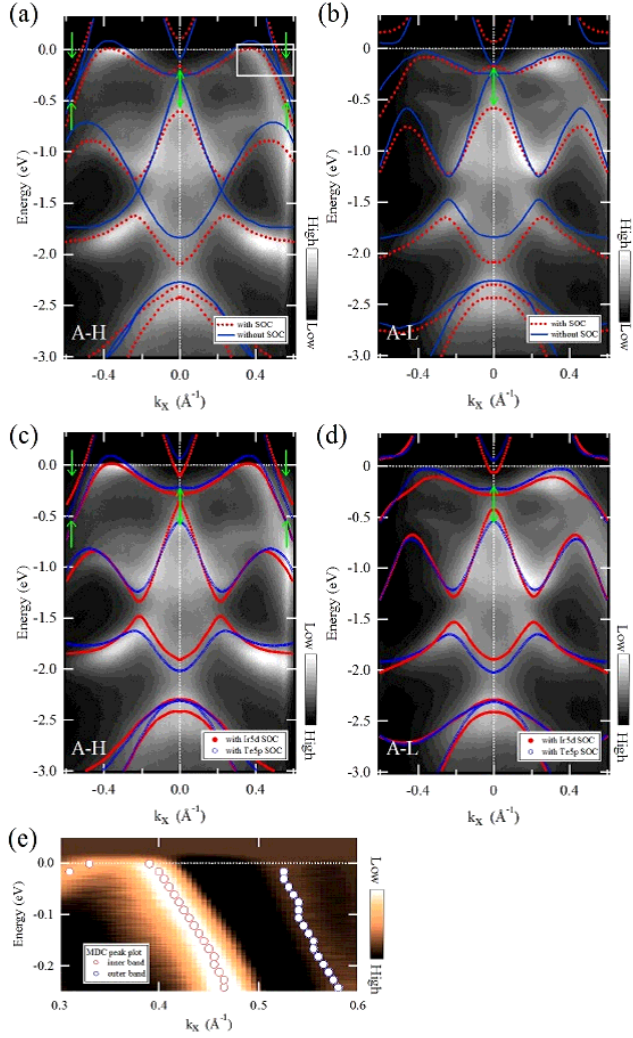}
\caption{(color online) 
ARPES spectra of Ir$_{0.95}$Pt$_{0.05}$Te$_2$
along the A-H direction (a) and along the A-L direction (b)
compared with the GGA calculations with and without
the spin-orbit interactions.
The red dots and light blue curves indicate the band dispersions
calculated with and without the spin-orbit interaction, respectively.
The arrows indicate spin-orbit splitting.
ARPES spectra of Ir$_{0.95}$Pt$_{0.05}$Te$_2$
along the A-H direction (c) and along the A-L direction (d)
compared with the GGA calculation with (without) the Te 5$p$ (Ir 5$d$) 
spin-orbit interaction and that without (with) the Te 5$p$ (Ir 5$d$).
The red closed circles and the blue open circles indicate the band dispersions calculated 
with the Ir 5$d$ and Te 5$p$ spin-orbit interactions, respectively.
The arrows indicate spin-orbit splitting. 
(e) Second derivative plot of momentum distribution curves(MDC) for the region indicated 
by the box in panel (a). The open circles indicate the peak position obtained from MDC.  
}
\end{figure}

\end{document}